\documentclass[twocolumn,english,aps,pra,superscriptaddress]{revtex4}
\usepackage{amsmath}
\usepackage{amssymb}

\usepackage{graphicx}

\makeatother

\usepackage{babel}
\begin{document}

\title{Multi-Particle Quantum Szilard Engine with Optimal Cycles Assisted
by a Maxwell's Demon}

\author{C. Y. Cai
}

\affiliation{Institute of Theoretical Physics, Chinese Academy of Sciences, Beijing,
100190, China}

\author{H. Dong
}

\affiliation{Institute of Theoretical Physics, Chinese Academy of Sciences, Beijing,
100190, China}

\author{C. P. Sun
}

\email{suncp@itp.ac.cn}

\homepage{http://power.itp.ac.cn/suncp/index.html}

\affiliation{Institute of Theoretical Physics, Chinese Academy of Sciences, Beijing,
100190, China}
\begin{abstract}
We present a complete-quantum description of multi-particle Szilard
engine which consists of a working substance and a Maxwell's demon.
The demon is modeled as a multi-level quantum system with specific
quantum control and the working substance consists of identical particles
obeying Bose-Einstein or Fermi-Dirac statistics. In this description,
a reversible scheme to erase the demon's memory by a lower temperature
heat bath is used. We demonstrate that (1) the quantum control of
the demon can be optimized for single-particle Szilard engine so that
the efficiency of the demon-assisted thermodynamic cycle could reach
the Carnot cycle's efficiency; (2) the low-temperature behavior of
the working substance is very sensitive to the quantum statistics
of the particles and the insertion position of the partition. 
\end{abstract}


\maketitle

\section{\label{sec1}Introduction}

Maxwel introduced, in 1871, a notorious being, known as Maxwell's
demon nowadays, to discuss the ``limitations of the second law
of thermodynamics''\cite{MD_book}. Such a demon distinguishes
the velocities of the gas particles and controls a tiny door on a
partition of the gas container to create a temperature difference,
which breaks the Clausius statement of the Second Law of Thermodynamics
(SLoT). To reveal the essence of the Maxwell's demon, Leo Szilard
proposed a single-particle heat engine \cite{Szilard1929}, named
as Szilard heat engine (SHE). The demon in SHE distinguishes the positions
of the particles after the partition has been inserted. With the help
of the demon, SHE can absorb heat from a single heat resource and
convert it into work without apperant evoking other changes in the
cycle. Szilard pointed out that the SLoT was no longer violated if
one considered the entropy increase during the measurement. Brillouin
generalized the Szilard's argument and identified the thermodynamic
entropy with the informational entropy firstly \cite{Brillouin1951}.

However, the measurement could be carried out without any change in
entropy \cite{Bennett1982}. Acutally, it was realized that a logically
irreversible process must be ``accompanied by dissipative effects''
\cite{Landauer1961} in the physical realization of the information
processing, which is known as Landauer's erasure principle. Bennett
used this point of view in the study of the Maxwell's demon paradox
and pointed out that the erasure of the demon's memory instead of
the measurement was logically irreversible and thus must be accompanied
by dissipative effects \cite{Bennett1982}. With these observations,
the conventional cycle presented by Szilard is indeed not a thermodynamic
cycle because the demon's memory has not been erased to complete the
cycle. The SLoT will be saved if one considers the erasure process
to finish the cycle of the demon. As people believe the essence of
information should be discussed in the framework of quantum mechanics,
various quantum versions of SHE have been proposed with different
views about quantum measurements. One proposal is semi-classical \cite{Plenio and Vitelli 2001,Ueda2010}.
The working substance in this proposal is quantum mechanic while the
demon is considered as a classical controller whose role is to extract
information through measurement and control the system. The paradox
of Maxwell's demon was solved by arguing Landauer's erasure principle.
However, the final solution should include the MD in the cycle and
treat also the MD in a quantum fasion\cite{Quan2006,Lloyd1997}. For
SHE, it is also proved the existence of MD will not violate the SLoT
in Ref. \cite{Dong2011}, where MD is modeled as a two-level system.

The focus of the study in both classical and quantum mechanical frameworks
is the erasure process, which is crucial to solving the Maxwell's
demon paradox. In an odinary way, the demon and the working substance
are in contact with the same heat bath. After erasing the demon by
applying some work, one will find that SHE can not extract work in
a cycle at all and the SLoT is not violated. However, a more general
erasure should be done with a lower-temperature heat bath
which is also called a heat sink. In this suituation,
we turn the SHE into a thermal dynamic cycle, where the non-violation
of SLoT can be proved by illustrating non-exceeding of Carnot's efficiency.
It is realized that the effective temperature of the MD's initial state actually
charasticrizes the error in the control of the heat engine\cite{Quan2006,Dong2011}.
Our previous work in Ref.~\cite{Dong2011} emphasized the functions
of the demon with errors in the study of the single-particle SHE.
But the erasure schemes in Refs.~\cite{Quan2006,Dong2011} are irreversible.
Therefore, the efficiencies of the heat engines in these papers can
not reach the Carnot cycle's efficiency. One purpose of the present
paper is to establish an optimal scheme of the thermodynamic cycle
with a reversible erasure process, assisted by demon. It is shown
that the partition-removing process is not always reversible, which
leads to the lower efficiency. We aso find the existence of the optimal
expansion position to improve the efficiency of the single particle
SHE to the Carnot cycle's efficiency. The other purpose of this paper
is to reveal the role of the quantum statistical properties of the
working substance. We generalize our previous works about demon-assisted
quantum heat engine by using a multi-particle working substance, which
is the ideal Bose or Fermi gases, and find that the ratio of the
work extracted to the working temperature has some discontinuous behavior,
and that discontinuous behavior is closely related to the degenerate-ground-state
phenomenon.

The paper is organized as follows: In Sec.~\ref{sec2} we describe
the model of quantum multi-particle SHE and the working scheme briefly.
In Sec.~\ref{sec3}, we study in details the five steps of the working
scheme: insertion, measurement, controlled expansion, removing and
erasure separately and calculate the work applied and heat transferred
in each step. In Sec.~\ref{sec4}, the efficiency of the engine is
evaluated. It is found that our heat engine's efficiency can not exceed
the Carnot cycle's efficiency. For single-particle SHE, we optimize
the scheme to make the efficiency of the engine reach the one of Carnot's.
In Sec.~\ref{sec5}, we discuss the behavior of the engine in low-temperature
regime and show the ratio of the work extrated to the working temperature
is closely related to the degenerate point (whose definition can be
seen in this section) and the particles' statistical properties. Conclusions
and remarks are given in Sec.~\ref{sec6}.

\begin{figure}[!htb]
 \includegraphics[width=8cm]{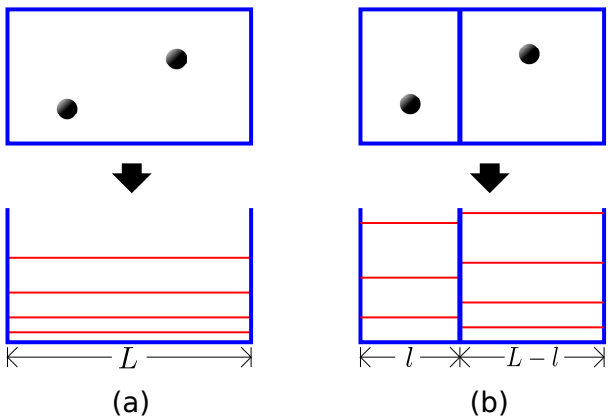} \caption{\label{chamber} (Color Online) Modeling the chamber of gas by a one-dimensional
infinite square well. (a) The chamber is modeled as a one-dimensional
infinite square well with width $L$ where the particles are confined.
The inner lines in the well represent the single-particle energy levels
of the well. (b) After the insertion, the chamber is split into two
chambers, which are modeled as two 1-D infinite square wells with
widths $l$ and $L-l$ respectively. The inner lines represent the
single-particle energy levels. }
\end{figure}

\begin{figure}[!htb]
 \includegraphics[width=8cm]{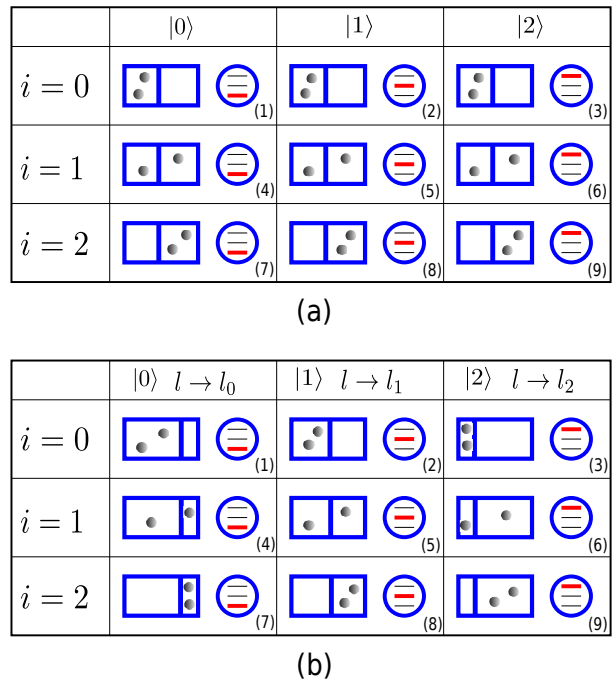} \caption{\label{expansion}
(Color Online) The states of the chamber and the demon
(a) before and (b) after the expansion process. (a) After the measurement,
the demon has revealed the number of the particles in the right compartment
$i$ and recorded this information in its memory. If the demon is
error-free, then its inner state (or memory) will be $\left\vert i\right\rangle $.
However, when the demon contains some errors, there would be some
mistakes in the demon's memory. For example, the sub-figure(2) shows
a situation when $i=0$ and the demon is in state $\left\vert 1\right\rangle $.
Thus, there are total $(N+1)^{2}$ situations after the measurement
for a system with $N$ particles. Here, the total $9$ situations
are listed for $N=2$. (d) After the measurement, the partition acts
as a piston under the control of the demon. If the demon is in the
state $\left\vert i\right\rangle $, the piston will move to a new
position $l_{i}$. Such a movement is called controlled expansion.
Here the results of the expansion for all $9$ situations are listed
for $N=2$.}
\end{figure}

\section{\label{sec2} Quantum Multi-Particle Szilard Engine}

The working substance is modeled as a collection of $N$ particles
confined in a chamber, which is described as a one-dimensional infinitely
deep square well with width $L$ illustrated in Fig.~\ref{chamber}(a).
Generally, the particles can satisfy any distribution, Bose distribution
or Fermi distribution. We calculate the canonical partition function
$Z_{n}(l)$ as follows, where the subscript $n$ denotes the number
of the particles in the well and $l$ represents the width of the
well. Its explicit expression is given in the Appendix \ref{appendix1}.
In the following calculations, we do not specify the concrete partition
functions in most cases. The results obtained are valid for both Bose
and Fermi distributions except for the case we claim specifically.

In the model, Maxwell's demon is described as a quantum system with
$N+1$ energy levels to record the complete information about how
many particles in each compartment after the insertion process. We
may use a demon with number of energy levels less than $N+1$, but
it can only record partial information of the working substance, which
decreases the work extract. We consider in the present paper only the
demon with $N+1$ levels, with $\left\vert i\right\rangle $ be its
$i$-th eigen-state and $\Delta_{i}$ be the energy of state $\left\vert i\right\rangle $.
The energy of $\left\vert 0\right\rangle $ is set to be the zero
point of energy, i.e., $\Delta_{0}=0$. For a demon without errors,
its initial state is a pure state, namely, $\rho_{d0}=\left\vert 0\right\rangle \left\langle 0\right\vert $.
In this paper, we study the demon with errors generally, whose initial
state can be written as 
\begin{equation}
\rho_{d0}=p_{0}^{(0)}\left\vert 0\right\rangle \left\langle 0\right\vert +p_{1}^{(0)}\left\vert 1\right\rangle \left\langle 1\right\vert +\cdots+p_{N}^{(0)}\left\vert N\right\rangle \left\langle N\right\vert .\label{ini_demon}
\end{equation}
 When $p_{i}^{(0)}$ vanish for all $i\not=0$, this demon returns
to the one without errors, where$\left\vert 0\right\rangle $ is the
standard state. For a demon with errors, a larger population $p_{0}^{(0)}$
in the state $\left\vert 0\right\rangle $ means a more efficient
demon \cite{Dong2011}.

The thermodynamic cycle consists the five steps. At the beginning,
the working substance is in equilibrium with the heat resource at
a higher temperature $T_{1}$, and the demon is initially in the state
$\rho_{d0}$ (see Eq.~(\ref{ini_demon})). In the first step, a partition
is inserted at the position $l$, which splits the initial well into
two with widths being $l$ and $L-l$ respectively (see Fig.~\ref{chamber}(b)).
After the insertion, there are $N+1$ possible situations and one
does not know which situation happens. Suppose $i$ to be the number
of the particles in the right compartment. Then the $N+1$ situations
correspond to $i=0,1,\cdots,N$. In the second step, the demon measures
the working substance and records this information in its memory.
For the error-free demon, it would be in the state $\left\vert i\right\rangle $
for the $i$-th situation after the measurement. However, the demon
with errors may record wrong information. For example, the working
substance is in 0-th situation while the demon is in the state $\left\vert 1\right\rangle $
as illustrated in Fig.~\ref{expansion}(a). In the third step, the
partition acts as a piston and expands under the control of the demon.
Specifically, if the demon is in the state $\left\vert i\right\rangle $,
it will move the piston to the position $l_{i}$ (see Fig.~\ref{expansion}(b)).
Actually, one can optimize the expansion position $l_{i}$ to get
the best efficiency. In the fourth step, the partition is removed
and the working substance returns to its initial state. After these
four steps, the working substance absorbs heat from the heat resource
and converts it into work. It seems to violate the SLoT. However,
the demon has not returned to its initial state to complete the cycle.
Thus, an additional step, step five, is needed to erase the memory
of the demon. The effect of this step saves the SLoT as pointed out
by Landauer's erasure principle.

\section{\label{sec3} Quantum Thermodynamic Cycle of Multi-Particle Szilard
Engine}

In this section, we analyze in detail the thermodynamic cycle of the quantum
multi-particle SHE. We calculate the heat absorbed and work applied
in each step.

\subsection*{\label{insertion}Step 1: Insertion}

In the first step, the partition is inserted isothermally, which means
the working substance is in contact with a heat resource and the process
is quasi-static. Before the insertion, the state of working substance
$\rho_{s0}$ is 
\begin{equation}
\rho_{N}(L)=\frac{1}{Z_{N}(L)}e^{-\beta_{1}H_{N}(L)},
\end{equation}
where $\beta_{1}=1/T_{1}$ is the inverse temperature and the Boltzmann
constant is set to be unit, i.e., $k_{B}=1$, and $H_{N}(L)$ represents
the Hamiltonian of $N$ particles confined in the infinite square
well with width $L$. $Z_{N}(L)$ is the corresponding partition function
and its explicit expression is presented in Appendix \ref{appendix1}.
After the insertion, the chamber is split into two parts with widths
$l$ and $L-l$ respectively and the substance is in an equilibrium
state as 
\begin{equation}
\rho_{s,\mathrm{ins}}=\sum_{i=0}^{N}P_{i}(l)\rho_{N-i}^{L}(l)\otimes\rho_{i}^{R}(L-l).
\end{equation}
 Here, $\rho_{N-i}^{L}(l)$ describes an equilibrium state of the
left chamber with similar meaning as those in $\rho_{N}(L)$. 
\begin{equation}
P_{i}(l)=\frac{Z_{N-i}(l)Z_{i}(L-l)}{\sum_{i=0}^{N}Z_{N-i}(l)Z_{i}(L-l)}
\end{equation}
 is the probability to find $i$ particles in the right chamber with
insertion position being $l$. As the process is isothermal, the work
applied in this step is the difference of the free energies after
and before the insertion, i.e., 
\begin{equation}
W_{\mathrm{ins}}=F_{\mathrm{s,ins}}-F_{s0},
\end{equation}
 and the heat absorbed is 
\begin{equation}
Q_{\mathrm{ins}}=T_{1}(S_{\mathrm{s,ins}}-S_{s0}).
\end{equation}
 The free energy and the entropy are given in terms of the partition
functions as 
\begin{eqnarray}
F_{s,\mathrm{ins}} & = & -T_{1}\ln\left(\sum_{i=0}^{N}Z_{N-i}(l)Z_{i}(L-l)\right),\\
F_{s0} & = & -T_{1}\ln Z_{N}(L),
\end{eqnarray}
 and 
\begin{eqnarray}
S_{s,\mathrm{ins}} & = & (1-\beta_{1}\frac{\partial}{\partial\beta_{1}})\ln\left(\sum_{i=0}^{N}Z_{N-i}(l)Z_{i}(L-l)\right),\\
S_{s0} & = & (1-\beta_{1}\frac{\partial}{\partial\beta_{1}})\ln Z_{N}(L),
\end{eqnarray}
 respectively. As shown in Ref.~\cite{Ueda2010}, in quantum mechanics
framework, the insertion work $W_{ins}$ is no longer zero. Moreover,
it was proved in Ref.~\cite{Dong2011} that $\lim_{T\rightarrow\infty}W_{ins}=\infty$.

\subsection*{\label{Measurement}Step 2: Measurement}

In the second step, the total system is isolated from the heat bath.
The demon finds out the number of the particles in the right compartment.
This measurement process aims at establishing the correlation between
the working substance and the demon. For the demon with no error,
the state of the demon after the measurement is $\left\vert i\right\rangle $
when there are $i$ particles in the right compartment. For the demon
with errors, its initial state is a mixed state. Thus, we should appoint
the final states of the demon for all situations. Let the final state
be $\left\vert f_{i}(j)\right\rangle $ when the demon is initially
in the state $\left\vert j\right\rangle $ and the number of the particles
in the right compartment is $i$. Mathematically, $f_{i}:j\mapsto f_{i}(j)$
is a map from the set $\{0,1,\cdots,N\}$ to itself. If the demon
is error-free, the final state will be $\left\vert i\right\rangle $
when there are $i$ particles in the right compartment, which leads
the first constraint of $f_{i}$, i.e., $f_{i}(0)=i$. With this notation,
the operator representing this measurement is 
\begin{equation}
U=\sum_{i=0}^{N}\left[\sum_{l_{i}}\left\vert \psi_{l_{i}}^{i}\right\rangle \left\langle \psi_{l_{i}}^{i}\right\vert \otimes\sum_{j=0}^{N}\left\vert f_{i}(j)\right\rangle \left\langle j\right\vert \right],
\end{equation}
 where $\left\vert \psi_{l_{i}}^{i}\right\rangle $ represents the
$l_{i}$-th eigen-state of the working substance when there are $i$
particles in the right compartment. For a physical operation, $U$
should be unitary (so that it is a quantum nondemolition pre-measurement),
which leads to another constraint of $f_{i}(j)$. That is, for a fixed
$i$, when $j$ runs over from $0$ to $N$, $f_{i}(j)$ should also
run over from $0$ to $N$. Thus, $\{f_{i}(j)\vert j=0,1,\cdots,N\}$
is a permutation of $\{0,1,\cdots,N\}$ which satisfies $f_{i}(0)=i$.
When $N=1$, there is only one function of such kind and the corresponding
operator $U$ represents a controlled-NOT operation. For $N>1$, such
function exists. One realization is $f_{i}(j)\equiv i+j\pmod{N+1}$.

After the measurement, the state of the total system becomes $\rho_{\mathrm{mea}}=U\rho_{s,\mathrm{ins}}\otimes\rho_{d0}U^{\dagger}$,
namely \begin{widetext} 
\begin{equation}
\rho_{\mathrm{mea}}=\sum_{j=0}^{N}\left\vert j\right\rangle \left\langle j\right\vert \otimes\sum_{i=0}^{N}P_{i}(l)p_{f_{i}^{-1}(j)}^{(0)}\rho_{N-i}^{L}(l)\otimes\rho_{i}^{R}(L-l),\label{state_aft_mea}
\end{equation}
 \end{widetext} where $f_{i}^{-1}:j\mapsto f_{i}^{-1}(j)$ is the
inverse of the map $f_{i}$. It follows from Eq.~(\ref{state_aft_mea})
that there is a possibility that the demon is in the state $\left\vert i\right\rangle $
while the number of particles in the right compartment is not $i$.
That is exactly what we mean `error' within the demon. Due to the
isolation from the heat bath, there is no heat transferred, $Q_{\mathrm{mea}}=0$,
and the work applied is just the energy change during the measurement:
\begin{equation}
W=\sum_{j=0}^{N}\left[\left(\sum_{i=0}^{N}P_{i}(l)p_{f_{i}^{-1}(j)}^{(0)}\right)-p_{j}^{(0)}\right]\Delta_{j},
\end{equation}
 which compensates for the energy difference between the initial state
and the final state of the demon during the measurement. Therefore,
if all the energy levels of the demon are degenerate, one will not
apply any work to perform the measurement. Thus the work applied may
not always be non-zero \cite{Bennett1982}.

\subsection*{\label{Expansion}Step 3: Controlled Expansion}

In this step, the partition in the chamber acts as a piston and will
be moved to the position according to the memory of the demon. Specifically,
if the state of the demon is $\left\vert i\right\rangle $, the finial
position will be $l_{i}$, which is called expansion position. This
expansion process is slow enough and the working substance is in contact
with the heat resource during the total expansion process. Thus the
state of the total system after the controlled expansion is 
\begin{equation}
\rho_{\exp}=\sum_{j=0}^{N}\left\vert j\right\rangle \left\langle j\right\vert \otimes\sum_{i=0}^{N}P_{i}(l)p_{f_{i}^{-1}(j)}^{(0)}\rho_{N-i}^{L}(l_{j})\otimes\rho_{i}^{R}(L-l_{j}).\label{state_aft_exp}
\end{equation}

However, the `impenetrability' of the piston does not make this process
an isothermal process since the transition between two states with
different numbers of the particles in the right compartment is forbidden.
Thus we should deal with this expansion process for each situation
separately. For the situation when the demon is in state $\left\vert j\right\rangle $
and there are $i$ particles in the right compartment, the work applied
$W_{ij}$ is 
\[
W_{ij}=-T_{1}\ln Z_{N-i}(l_{j})Z_{i}(L-l_{j})+T_{1}\ln Z_{N-i}(l)Z_{i}(L-l),
\]
 and the heat absorbed $Q_{ij}$ is 
\[
Q_{ij}=T_{1}(S(\rho_{N-i}^{L}(l_{j})\otimes\rho_{i}^{R}(L-l_{j}))-S(\rho_{N-i}^{L}(l)\otimes\rho_{i}^{R}(L-l))).
\]
 The probability of this situation is $p_{ij}=P_{i}(l)p_{f_{i}^{-1}(j)}^{(0)}$.
Thus, the total work applied and the total heat absorbed are: 
\begin{eqnarray}
W_{\mathrm{exp}} & = & \sum_{j=0}^{N}\sum_{i=0}^{N}p_{ij}W_{ij},\\
Q_{\mathrm{exp}} & = & \sum_{j=0}^{N}\sum_{i=0}^{N}p_{ij}Q_{ij},
\end{eqnarray}
 respectively. One can prove that even if the controlled expansion
is not isothermal, the heat absorbed also satisfies the relationship
$Q=T\Delta S$, where $\Delta S$ is the entropy change of the total
system. Since the states with different numbers of the particles in the
right compartment are orthogonal to each other, the entropies of the
total system before and after the controlled expansion are 
\begin{eqnarray}
S_{\mathrm{mea}} & = & H(\{p_{ij}\})+\sum_{j=0}^{N}\sum_{i=0}^{N}p_{ij}S_{mea,ij},\\
S_{\mathrm{exp}} & = & H(\{p_{ij}\})+\sum_{j=0}^{N}\sum_{i=0}^{N}p_{ij}S_{exp,ij},
\end{eqnarray}
 respectively, where $H(\{p_{ij}\})=-\sum_{ij}p_{ij}\ln p_{ij}$.
Thus, we have the following relationship 
\begin{eqnarray}
T_{1}\Delta S & = & T_{1}(S_{\mathrm{exp}}-S_{\mathrm{mea}})\nonumber \\
 & = & \sum_{j=0}^{N}\sum_{i=0}^{N}p_{ij}T_{1}(S_{exp,ij}-S_{mea,ij})\nonumber \\
 & = & \sum_{j=0}^{N}\sum_{i=0}^{N}p_{ij}Q_{ij}=Q_{\exp}.
\end{eqnarray}

\begin{figure}[!htb]
\includegraphics[width=8cm]{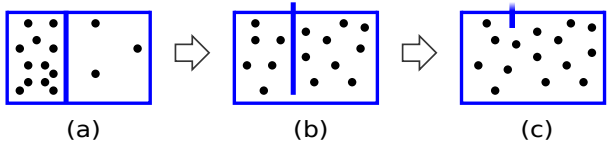} \caption{\label{fig_rev}
(Color Online) The removing process. (a) Before the removing,
the state of the working substance may not be an equilibrium state
because of the `impenetrability' of the partition (see the sub-section
\ref{Expansion}). (b) When the partition is lifted a little, it is
no longer impenetrable and the particles can fly between the two sides
of the chamber freely to reach the equilibrium state. (c) The partition
keeps being lifted and finally be removed isothermally. }
\end{figure}

\subsection*{\label{Removing}Step 4: Removing}

After the controlled expansion, the partition will be removed slowly
while the working substance is in contact with the heat resource.
Because the state in Eq.~(\ref{state_aft_exp}) before removing is
not a thermal equilibrium state (it is because the controlled expansion
is not an isothermal process, see the last subsection), one should
carefully deal with this step. Once lifted a little, the partition
is no longer `impenetrable' and the particles can fly between the
two sides of the chamber freely to reach an equilibrium state. Thus
the removing process consists of two sub-steps. After the first sub-step
finished, the partition has been lifted to create a small slit with
width $d_{p}$ through which the particles can fly from one side to
the other. Whatever small the width $d_{p}$ of the slit is, the working
substance can reach its equilibrium state. Thus, we can let the width
of the slit tend to zero, i.e., $d_{p}\rightarrow0$, which makes
this sub-step a thermalization process of the working substance while
the partition remains still. Thereafter the state of the system after
the first sub-step becomes \begin{widetext} 
\begin{equation}
\rho_{\mathrm{rev}}^{^{\prime}}=\sum_{j=0}^{N}p_{j}^{(1)}\left\vert j\right\rangle \left\langle j\right\vert \otimes\sum_{i=0}^{N}P_{i}(l_{j})\rho_{N-i}^{L}(l_{j})\otimes\rho_{i}^{R}(L-l_{j}),
\end{equation}
 where $p_{j}^{(1)}=\sum_{i=0}^{N}P_{i}(l)p_{f_{i}^{-1}(j)}$. During
this process, there is no work applied (since $d_{p}\rightarrow0$)
and the heat absorbed is exactly the difference between the inner
energies of the total system before and after the first sub-step,
\begin{equation}
Q_{\mathrm{rev}}'=\sum_{j=0}^{N}\sum_{i=0}^{N}\left[p_{j}^{(1)}P_{i}(l_{j})-P_{i}(l)p_{f_{i}^{-1}(j)}^{(0)}\right]\left[U(\rho_{N-i}^{L}(l_{j}))+U(\rho_{i}^{R}(L-l_{j}))\right].
\end{equation}
After the thermalization, the working substance is in a thermal equilibrium
state, which makes the rest sub-step, the second sub-step, an isothermal
process. After the removing, the working substance returns to its
initial state and the correlation between the working substance and
the demon no longer exists. During the second sub-step, the work applied
and the heat absorbed are 
\begin{eqnarray}
W_{\mathrm{rev}} & = & T_{1}\left[-\ln Z_{N}(L)+\sum_{j=0}^{N}p_{j}^{(1)}\ln\left(\sum_{i=0}^{N}Z_{N-i}(l_{j})Z_{i}(L-l_{j})\right)\right], \\
Q_{\mathrm{rev}} & = & T_{1}\left[S(\rho_{s0})-\sum_{j=0}^{N}p_{j}^{(1)}S(\rho_{s,ins}(l_{j}))\right],
\end{eqnarray}
 \end{widetext} respectively, and the state of the total system after the
removing process is 
\begin{equation}
\rho_{\mathrm{rev}}=\sum_{j=0}^{N}p_{j}^{(1)}\left\vert j\right\rangle \left\langle j\right\vert \otimes\rho_{s0}.\label{state_aft_rev}
\end{equation}
 One can see that the state of the demon does not return to its initial
state, which does not make the above four steps a thermodynamic cycle.
Thus, an addition step, erasure, is needed.

\begin{figure}[!htb]
\includegraphics[width=8cm]{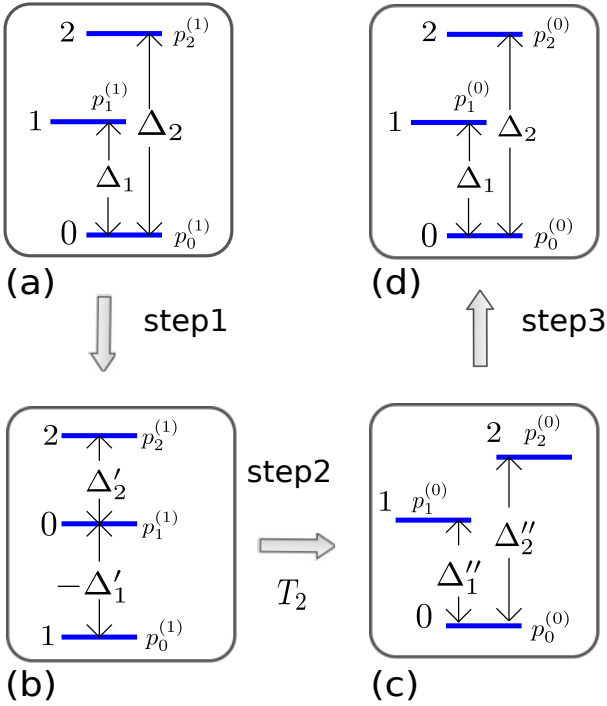} \caption{\label{fig_era} (Color Online)
Erasing the memory of a three-energy-level demon
reversibly. Suppose that the populations of the three
levels after the removing process the partition are $p_{0}^{(1)}=0.3$, $p_{1}^{(1)}=0.5$
and $p_{2}^{(1)}=0.2$ as in (a) and the initial state's populations
of the three levels are $p_{0}^{(0)}=0.7$, $p_{1}^{(0)}=0.2$ and
$p_{2}^{(0)}=0.1$ as in (d). In step 1 (a-b), the demon is isolated
from the heat bath and the energy level spacings are adjusted from
$\Delta_{i}$ to $\Delta_{i}^{\prime}$ for $i=1,2$. During this process,
the populations of the levels remain constant. The new level spacings
are chosen so that the demon has an effective temperature $T_{2}=1$
after the adjusting, which implies $\Delta_{1}^{\prime}=-0.5108$
and $\Delta_{2}^{\prime}=0.4055$. In (b), the minus sign in front
of $\Delta_{1}^{\prime}$ is due to the negative value of $\Delta_{1}^{\prime}$.
In step 2 (b-c), the demon is in contact with a real heat bath at
temperature $T_{2}$ and the energies of the levels are adjusted.
In this step, the populations of the levels change while its energy
spacings change. The goal of this step is to let the populations of
the levels return to be $p_{0}^{(0)}$, $p_{1}^{(0)}$ and $p_{2}^{(0)}$.
Thus the new level spacings are $\Delta_{1}^{\prime\prime}=1.2528$
and $\Delta_{2}^{\prime\prime}=1.9459$. In step 3 (c-d), the demon
is isolated from the heat bath again and its level spacings are adjusted
to return to its initial values. The isolation guarantees that the
populations of the levels are not changed. Thus the state of the demon
has returned to its initial state. }

\end{figure}

\subsection*{\label{Erasure}Step 5: Erasure}

In Eq.~(\ref{state_aft_rev}), the state of the demon after the removing
process is 
\begin{equation}
\rho_{d1}=\sum_{j=0}^{N}p_{j}^{(1)}\left\vert j\right\rangle \left\langle j\right\vert ,
\end{equation}
which is different from the initial state of the demon. As part of
the engine, the demon should be re-initialized to complete the cycle.
Because the entropy of $\rho_{d1}$ is different from the entropy
of $\rho_{d0}$, such a re-initialization process, or erasure process,
can not be realized by a unitary evolution and must be a logical irreversibility
process accompanied by dissipative effects \cite{Landauer1961}. A
natural way to complete the re-initialization process is to make use
of a thermalization process. It can be shown that the heat bath used
in the thermalization process is a heat sink instead of a heat resource.
Because the entropy of $\rho_{d1}$ is larger than the entropy of
$\rho_{d0}$, the heat must be transferred from the demon to the heat
bath. It seems that after a series of complex tasks, the demon needs
to be cooled down. According to thermodynamics, the best re-initialization
process is a reversible one which does not cost additional dissipative
work. Before going into the general discussion, we first consider
a three-level demon as an example to find out how to re-initialize
or erase the demon reversibly.

The three-level demon considered is shown in Fig.~\ref{fig_era},
where the three states are $\left\vert 0\right\rangle $, $\left\vert 1\right\rangle $
and $\left\vert 2\right\rangle $ respectively. The energy of the
state $\left\vert 0\right\rangle $ is set to be zero and the energy
of the state $\left\vert i\right\rangle $ is $\Delta_{i}$ for $i=1,2$.
For a demon in its initial state, the populations of the three levels
are $p_{0}^{(0)}$, $p_{1}^{(0)}$ and $p_{2}^{(0)}$ respectively
(see Fig.~\ref{fig_era}(d)). To be more concrete, we assume that
$p_{0}^{(0)}=0.7$ while $p_{1}^{(0)}=0.2$ and $p_{2}^{(0)}=0.1$.
Suppose that the populations become $p_{0}^{(1)}=0.3$, $p_{1}^{(1)}=0.5$
and $p_{2}^{(1)}=0.2$ respectively after the removing (see Fig.~\ref{fig_era}(a)).
In order to re-initialize it by using a reversible scheme by means
of a heat bath with $T_{2}=1$, we adjust the demon's energy level
spacings adiabatically (see Fig. \ref{fig_era} (a)-(b)). After being adjusted,
the level spacings are changed from $\Delta_{i}$ ($i=1,2$) to $\Delta_{i}^{\prime}$
(see Fig. \ref{fig_era}(b)). During this process, the population
of each state remains unchanged. Thus, the new level spacings satisfy
the following relationship to ensure that the effective temperature
of the demon is $T_{2}$, i.e., 
\begin{equation}
\frac{p_{i}^{(1)}}{p_{0}^{(1)}}=\exp\left(-\frac{\Delta_{i}^{\prime}}{T_{2}}\right),
\end{equation}
 for $i=1,2$, which implies $\Delta_{1}^{\prime}=-0.5108$ and $\Delta_{2}^{\prime}=0.4055$.
Next, the demon is kept in contact with the heat bath at temperature
$T_{2}$ without any undesired irreversible heat transfer, and we
adjust the energies of its levels isothermally (see Fig. \ref{fig_era}
(b)-(c)). In this process, the population of each level changes as
the level spacings change and the final populations are adjusted to
be $p_{0}^{(0)}$, $p_{1}^{(0)}$ and $p_{2}^{(0)}$ respectively,
which are the initial populations of the demon. Therefore the new
level spacings $\Delta_{i}^{\prime\prime}$ ($i=1,2$) satisfy the
relationship 
\begin{equation}
\frac{p_{i}^{(0)}}{p_{0}^{(0)}}=\exp\left(-\frac{\Delta_{i}^{\prime\prime}}{T_{2}}\right),
\end{equation}
 which implies $\Delta_{1}^{\prime\prime}=1.2528$ and $\Delta_{2}^{\prime\prime}=1.9459$
(see Fig. \ref{fig_era}(c)). After that, the population of each state
has returned to its initial value while the level spacings do not.
Thus we need another adiabatical adjusting process as in the first
step to re-initialize the level spacings (see Fig. \ref{fig_era}
(c)-(d)). These three sub-steps complete the re-initialization process
and are all reversible, which makes the total re-initialization process
reversible.

The above example inspires us to construct the reversible re-initialization
scheme in a more general case. In the following discussion, a detailed
scheme of the re-initialization process is presented. In the first
sub-step, we adjust the energy level of the state $\left\vert j\right\rangle $
for $j\neq0$ adiabatically while the energy of $\left\vert 0\right\rangle $
is kept fixed. The corresponding new state is denoted as $\left\vert j'\right\rangle $.
During this sub-step, the demon is isolated from the heat sink. The
target is to let the demon's effective temperature be $T_{2}$, which
is the temperature of the heat sink. Thus, the new level spacing $\Delta_{j}'$
satisfies 
\begin{equation}
\exp(-\beta_{2}\Delta_{j}')=\frac{\sum_{i=0}^{N}P_{i}(l)p_{f_{i}^{-1}(j)}}{\sum_{i=0}^{N}P_{i}(l)p_{f_{i}^{-1}(0)}},\label{new_energy}
\end{equation}
 and the demon after the first sub-step is in the state 
\begin{equation}
\rho_{d,\mathrm{era}1}=\sum_{j=0}^{N}\left(\sum_{i=0}^{N}P_{i}(l)p_{f_{i}^{-1}(j)}\right)\left\vert j'\right\rangle \left\langle j'\right\vert .
\end{equation}
 It should be noticed that Eq.~(\ref{new_energy}) always has a solution
except for some extreme conditions, for example, $\sum_{i=0}^{N}P_{i}(l)p_{f_{i}^{-1}(j)}=0$
or $T_{2}=0$. If these extreme conditions happen, we will use some
tricks to deal with it. In the case of $T_{2}=0$, we will introduce
a small positive quantity $\varepsilon$ and let $T_{2}^{\prime}$
be $T_{2}+\varepsilon$. Next we will use $T_{2}^{\prime}$ instead
of $T_{2}$ to calculate the parameters of the scheme such as $\Delta_{i}^{\prime}$
and $\Delta_{i}^{\prime\prime}$. Such a scheme is not reversible.
However, if $\varepsilon$ tends to zero, the scheme will tend to
behave as a reversible scheme. In the following, we assume that Eq.~(\ref{new_energy})
does have solution. In this sub-step, there is no heat transfer and
the work applied is 
\begin{equation}
W_{\mathrm{ini},1}=\sum_{j=1}^{N}\left(\sum_{i=0}^{N}P_{i}(l)p_{f_{i}^{-1}(j)}\right)(\Delta_{j}'-\Delta_{j}).
\end{equation}

In the second sub-step, demon is in contact with the heat sink and
the $j$-th level's energy is adjusted from $\Delta_{j}'$ to $\Delta_{j}''$
isothermally, where $\Delta_{j}''$ satisfies 
\begin{equation}
p_{i}=\frac{e^{-\beta_{2}\Delta_{i}''}}{\sum_{i}e^{-\beta_{2}\Delta_{i}''}},
\end{equation}
 and $\beta_{2}$ is the inverse temperature of the heat sink. Thus,
the state of the demon after this sub-step is 
\begin{equation}
\rho_{d,\mathrm{era}2}=\sum_{j=0}^{N}p_{j}\left\vert j''\right\rangle \left\langle j''\right\vert ,
\end{equation}
 where $\left\vert j''\right\rangle $ represents the state corresponding
to $\left\vert j'\right\rangle $ after the adjusting. The work applied
and the heat transferred during this process are 
\begin{eqnarray}
W_{\mathrm{ini},2}=-\frac{1}{\beta_{2}}\sum_{j=1}^{N}\left[\ln(1+e^{-\beta_{2}\Delta_{j}'})-\ln(1+e^{-\beta_{2}\Delta_{j}''})\right],\\
Q_{\mathrm{ini}}=T_{2}\left[H\left(\{p_{i}\}\right)-H\left(\left\{ \sum_{i=0}^{N}P_{i}(l)p_{f_{i}^{-1}(j)}\right\} \right)\right],
\end{eqnarray}
 respectively. Here $H(\{p_{i}\})=-\sum_{i}p_{i}\ln p_{i}$ represents
the Shannon entropy of the probability distribution $\{p_{i}\}$.

In the last sub-step, the demon is isolated from the heat sink again
and the energies of all of its levels are adjusted to its initial
values. Thus the final state of the demon is $\sum_{j}p_{j}\left\vert j\right\rangle \left\langle j\right\vert $,
which is indeed its initial state. During this sub-step,
no heat transfers and the work applied is 
\begin{equation}
W_{\mathrm{ini},3}=\sum_{j=1}^{N}p_{j}(\Delta_{j}-\Delta_{j}'').
\end{equation}

These three sub-steps together make up the re-initialization scheme.
Such a scheme involves adjusting the energies of the demon's levels,
which is not easily realized. In a concrete experiment, if the quantum
system that functions as the Maxwell's demon has unchangeable energy
levels, we will only use an irreversible scheme as in Refs.~\cite{Quan2006,Dong2011}.
As we know in thermodynamics, such an irreversible re-initialization
scheme reduces the efficiency of the heat engine because of the dissipative
effects. Thus, when it comes to the Maxwell's demon paradox, one needs
a reversible re-initialization scheme without dissipative effects
to get a good physical picture.

\section{\label{sec4} Efficiency of Szilard Engine}

Combining the results obtained in the above section, one finally gets
the total heat transferred and the work extracted in the thermodynamic
cycle:

1. The heat absorbed from the heat resource is 
\begin{eqnarray}
Q_{1} & = & Q_{\mathrm{ins}}+Q_{\mathrm{exp}}+Q_{\mathrm{rev}}'+Q_{\mathrm{rev}}\label{Q1}\\
 & = & T_{1}\left[H\left(\{P_{i}(l)\}\right)+\sum_{j=0}^{N}\sum_{i=0}^{N}P_{i}(l)p_{f_{i}^{-1}(j)}\ln P_{i}(l_{j})\right].\nonumber 
\end{eqnarray}

2. The heat released to the heat sink is 
\begin{eqnarray}
Q_{2} & = & -Q_{\mathrm{ini}}\label{Q2}\\
 & = & T_{2}\left[H(\{\sum_{i=0}^{N}P_{i}(l)p_{f_{i}^{-1}(j)}\})-H\left(\{p_{i}\}\right)\right].\nonumber 
\end{eqnarray}

3. The work extracted is 
\begin{eqnarray}
W & = & -W_{\mathrm{ins}}-W_{\mathrm{mea}}-W_{\mathrm{exp}}-W_{\mathrm{rev}}\nonumber \\
 &  & -W_{\mathrm{ini},1}-W_{\mathrm{ini},2}-W_{\mathrm{ini},3},\label{W}
\end{eqnarray}
 which can be checked to satisfy the first law of thermodynamics,
i.e., $W=Q_{1}-Q_{2}$.

Firstly, it is emphasized that the result is general without referring
to the statistical properties of the working substance, since $P_{i}(l)$
has contained all the information of the distribution. Thus Eq.~(\ref{Q1}),
Eq.~(\ref{Q2}) and Eq.~(\ref{W}) are correct both for the Bose
system and the Fermi system. Secondly, one can optimize the expansion
position $l_{j}$ to maximize the efficiency of SHE. Consider a two-particle
SHE as an example, in which the demon is of three-level. The function
$f_{i}(j)$ we choose is $f_{i}(j)\equiv i+j\pmod{3}$. Thus the heat
absorbed from the heat resource is \begin{widetext} 
\begin{eqnarray}
Q_{1} & =T_{1}H(\{P_{i}(l)\}) & +P_{0}(l)p_{0}\ln P_{0}(l_{0})+P_{1}(l)p_{2}\ln P_{1}(l_{0})+P_{2}(l)p_{1}\ln P_{2}(l_{0})\nonumber \\
 &  & +P_{0}(l)p_{1}\ln P_{0}(l_{1})+P_{1}(l)p_{0}\ln P_{1}(l_{1})+P_{2}(l)p_{2}\ln P_{2}(l_{1})\nonumber \\
 &  & +P_{0}(l)p_{2}\ln P_{0}(l_{2})+P_{1}(l)p_{1}\ln P_{1}(l_{2})+P_{2}(l)p_{0}\ln P_{2}(l_{2}).\label{example_eff}
\end{eqnarray}
 Thereafter, the optimized expansion position $l_{i}$ should satisfy
\begin{equation}
\frac{\partial Q_{1}}{\partial l_{0}}=\frac{\partial Q_{1}}{\partial l_{1}}=\frac{\partial Q_{1}}{\partial l_{2}}=0.
\end{equation}
 Working out the above equation, one will get the optimized expansion
positions $l_{0\mathrm{max}}$, $l_{1\mathrm{max}}$ and $l_{2\mathrm{max}}$.

Thirdly, an upper limit of the heat absorbed from the heat resource
$Q_{1}$ in Eq.~(\ref{Q1}) is estimated. Use the method presented
in the Appendix \ref{appendix2}, one gets 
\begin{eqnarray}
Q_{1} & \leqslant & T_{1}\left[H\left(\{P_{i}(l)\}\right)+\sum_{i=0}^{N}\sum_{j=0}^{N}P_{i}(l)p_{f_{i}^{-1}(j)}\ln\frac{P_{i}(l)p_{f_{i}^{-1}(j)}}{\sum_{i=0}^{N}P_{i}(l)p_{f_{i}^{-1}(j)}}\right]\nonumber \\
 & = & T_{1}\left[H\left(\{\sum_{i=0}^{N}P_{i}(l)p_{f_{i}^{-1}(j)}\}\right)-H\left(\{p_{i}\}\right)\right].
\end{eqnarray}
 \end{widetext} Therefore the efficiency of the SHE has the upper
limit: 
\begin{eqnarray}
\eta & = & 1-\frac{Q_{2}}{Q_{1}}\nonumber \\
 & \leqslant & 1-\frac{T_{2}}{T_{1}}.
\end{eqnarray}
 Thus, it is clear that the efficiency of the SHE can not exceed the
Carnot cycle's efficiency when the erasure process or re-initialization
process is considered, which saves the SLoT. It can be seen more clearly
when $N=1$, which represents a single-particle SHE. For a single-particle
SHE, there is only one realization of $f_{i}(j)$, and the corresponding
measurement process is indeed a controlled-NOT operation as in Ref.~\cite{Quan2006,Dong2011},
i.e., $f_{0}(j)=j$ for $j=0,1$ while $f_{1}(1)=0$ and $f_{1}(0)=1$.
The formulas Eq.~(\ref{Q1}) and Eq.~(\ref{Q2}) are reduced to
\[
Q_{1}=T_{1}\left[\begin{array}{c}
H(P_{0}(l),P_{1}(l))\\
+P_{0}(l)p_{0}\ln P_{0}(l_{0})+P_{1}(l)p_{1}\ln P_{1}(l_{0})\\
+P_{0}(l)p_{1}\ln P_{0}(l_{1})+P_{1}(l)p_{0}\ln P_{1}(l_{1})
\end{array}\right],\eqno(\ref{Q1}')
\]
 and 
\[
Q_{2}=T_{2}\left[\begin{array}{c}
H(P_{0}(l)p_{1}+P_{1}(l)p_{0},P_{0}(l)p_{0}+P_{1}(l)p_{1})\\
-H(p_{0},p_{1})
\end{array}\right]\eqno(\ref{Q2}')
\]
 respectively. It is easy to maximize $Q_{1}$ to find that
$l_{0\mathrm{max}}$ and $l_{1\mathrm{max}}$ satisfy 
\begin{eqnarray}
P_{0}(l_{0\mathrm{max}}) & = & \frac{P_{0}(l)p_{0}}{P_{0}(l)p_{0}+P_{1}(l)p_{1}},\\
P_{0}(l_{1\mathrm{max}}) & = & \frac{P_{0}(l)p_{1}}{P_{0}(l)p_{1}+P_{1}(l)p_{0}},
\end{eqnarray}
 and the maximum of $Q_{1}$ is 
\begin{equation}
Q_{1\mathrm{max}}=T_{1}\left[\begin{array}{c}
H(P_{0}(l)p_{0}+P_{1}(l)p_{1},P_{0}(l)p_{1}+P_{1}(l)p_{0})\\
-H(p_{0},p_{1})
\end{array}\right].\label{Q1MAX}
\end{equation}
 Thus, it is found that $Q_{1\mathrm{max}}$ is always positive whatever
the insertion position is. Combining Eq.~(\ref{Q1MAX}) and Eq.~(\ref{Q2}),
one gets the best efficiency as 
\begin{eqnarray}
\eta_{\mathrm{max}} & = & 1-\frac{Q_{2}}{Q_{1\mathrm{max}}}\\
 & = & 1-\frac{T_{2}}{T_{1}}.
\end{eqnarray}
 Then it is concluded that for a single-particle SHE, one can always
choose a good expansion position to let the SHE reach the Carnot cycle's
efficiency even if the demon contains some errors. Recall that in
the step 4 of the thermodynamic cycle, the removing process, one needs
to thermalization the working substance firstly. However, for a single-particle
SHE, it is found that the expansion position can always be chosen
that the thermalization process is not needed. Thus, a single-particle
SHE can always be a reversible heat engine with two heat baths, and
its efficiency reaches the Carnot cycle's efficiency. On the other
hand, for a multi-particle SHE, the thermalization process is always
needed in the removing process, which makes the heat engine irreversible.
Thus, a multi-particle SHE's efficiency can not always reach Carnot cycle's
efficiency.

\begin{figure}[!htb]
\begin{minipage}[c]{0.25\textwidth}%
 \includegraphics[width=1\textwidth]{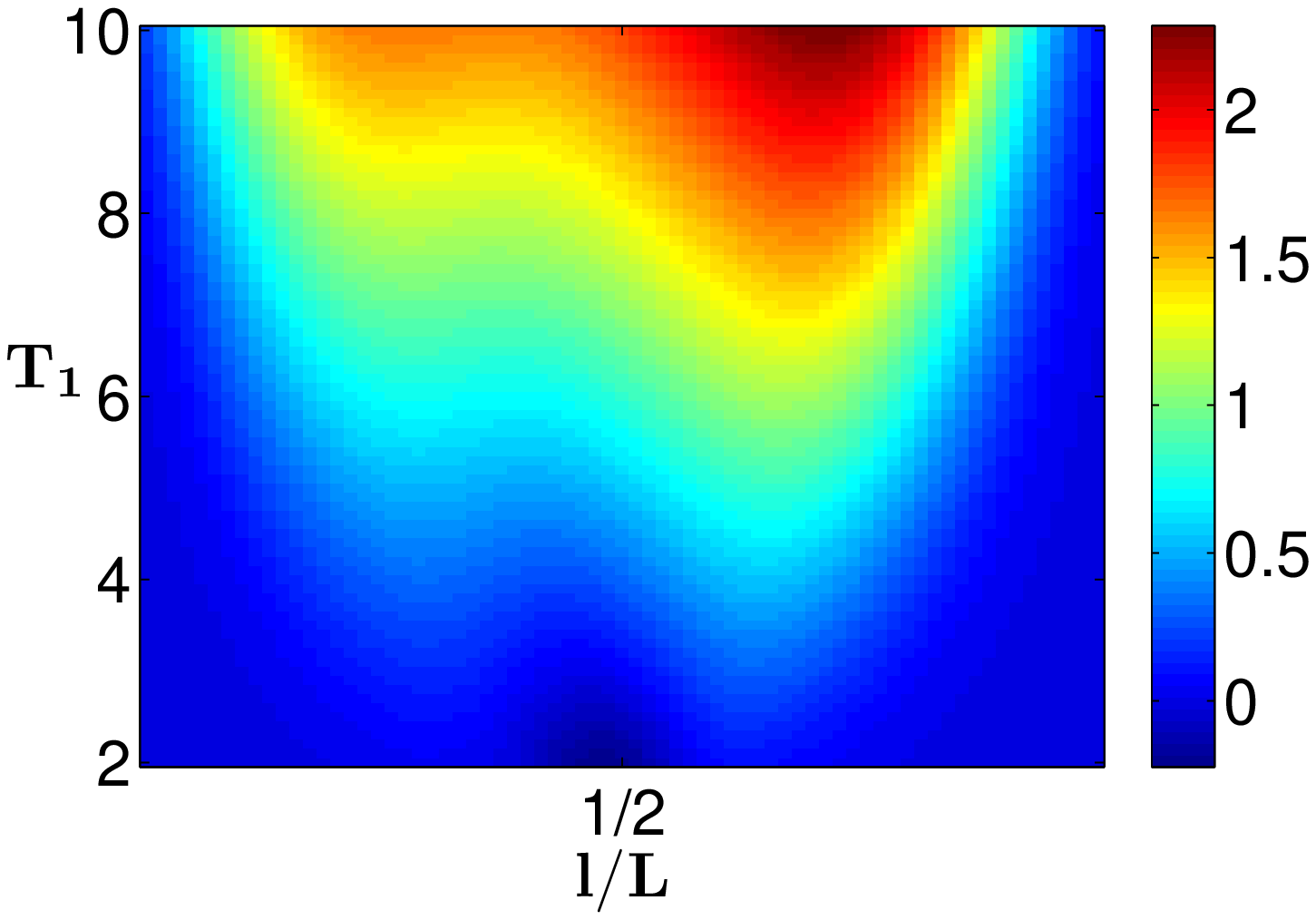} %
\end{minipage}%
\begin{minipage}[c]{0.25\textwidth}%
 \includegraphics[width=1\textwidth]{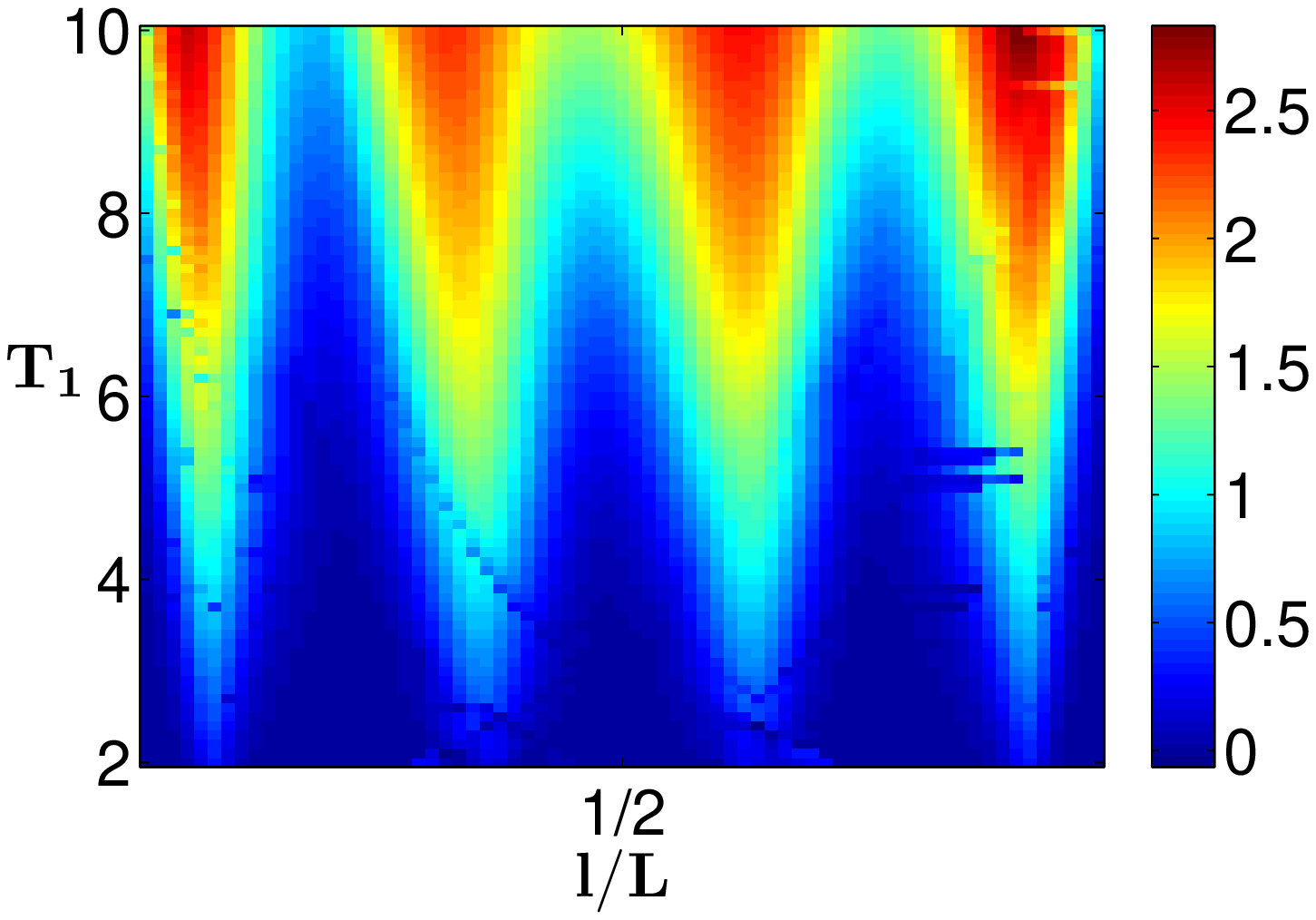} %
\end{minipage}\\
\begin{minipage}[c]{0.25\textwidth}%
 (1) Bose Case %
\end{minipage}%
\begin{minipage}[c]{0.25\textwidth}%
 (2) Fermi Case %
\end{minipage}\\
\begin{minipage}[c]{0.5\textwidth}%
 \caption{\label{bvf} (Color Online) Work extracted $W$ as the function of
insertion position $l/L$ and the working temperature $T_{1}$. (1)
Bose case with $N=4$. (2) Fermi case with $N=4$.}
\end{minipage}
\end{figure}

At the end of this section, a concrete example is presented to show
the difference between the Bose system and the Fermi system. In this
example, a four-particle system is considered as the working substance.
The demon we chosen is a five-level quantum system with some errors.
Its initial state is 
\begin{eqnarray}
\rho_{d0} & = & 0.7000\left\vert 0\right\rangle \left\langle 0\right\vert +0.2100\left\vert 1\right\rangle \left\langle 1\right\vert \nonumber \\
 & + & 0.0630\left\vert 2\right\rangle \left\langle 2\right\vert +0.0189\left\vert 3\right\rangle \left\langle 3\right\vert \nonumber \\
 & + & 0.0081\left\vert 4\right\rangle \left\langle 4\right\vert .
\end{eqnarray}
 The temperature of the heat sink is fixed that $T_{2}=1$. Denote
the higher temperature and the insertion position as $T_{1}$ and
$l$ respectively. Then the work extracted by the heat engine during
a thermodynamic cycle can be numerically calculated for both the Bose
case and the Fermi case. The results are shown in Fig.~\ref{bvf}.
As one can see, the patterns in Fig.~\ref{bvf} for different distributions
are quite different. In fact, in the next section, it is found that
the difference between the Bose case and the Fermi case is most significant
in low-temperature regime.

\section{\label{sec5} Low-Temperature Behavior with Different Quantum Statistics}

In this section, the behavior of a multi-particle SHE when the working
temperature $T_{1}$ is very low is studied. In Sec.~\ref{sec4},
a necessary condition for the heat engine to extract positive work
is found that $T_{1}>T_{2}$. Thus, if $T_{1}$ considered is very
small, the temperature of the heat sink will be $T_{2}=0$. The demon
in this case is free of errors because it is natural to get a pure
initial state if the demon is re-initialized by a zero-temperature
heat sink. The motivation to concern the demon without error is to
see the difference between the Bose system and Fermi system more clearly
without the effect of the demon's errors. As mentioned in the step
5 of Sec.~\ref{sec3}, it is less likely for one to re-initialize
such a demon by using a reversible scheme; however, one can use an
asymptotic scheme to re-initialize the demon.

\begin{figure}[!htb]
\includegraphics[width=8cm]{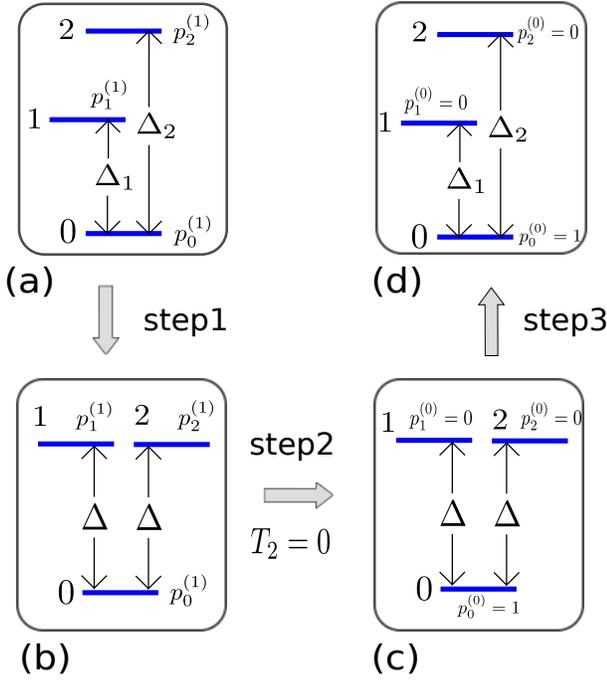} \caption{\label{fig_era1} (Color Online) Asymptotic erasure scheme for $T_{2}=0$. }

\end{figure}

Now, we describe the asymptotic scheme in details firstly. Similar
to the re-initialization scheme in Sec.~\ref{sec3}, the asymptotic
scheme also contains three sub-steps. The three sub-steps are illustrated
in Fig.~\ref{fig_era1}. In the first sub-step, the energies of the
levels $\left\vert i\right\rangle $ for $i\neq0$ are adjusted from
$\Delta_{i}$ to a very small value $\Delta(\neq0)$ adiabatically.
During this sub-step, there is no heat transferred. In the second
sub-step, the demon is in contact with the heat sink at temperature
$T_{2}=0$ until it is in equilibrium with the heat sink. During this
sub-step, there is no work applied and the finial state of the demon
is a pure state for its temperature is zero when $\Delta\neq0$. In
the third sub-step, the energies of the demon's levels are re-adjusted
adiabatically to its initial value. Thus, these three sub-steps complete
the re-initialization process of the demon. If we use this scheme,
$Q_{1}$ in Eq.~(\ref{Q1}) will not change and $Q_{2}$ in Eq.~(\ref{Q2})
will be revised to be the heat transferred in the second sub-step.
Due to the irreversibility of the second sub-step, this scheme is
not reversible. Nevertheless, if one lets $\Delta$ tend to $0$,
$Q_{2}$ will tend to $0$, which makes the scheme reversible. Thus,
Eq.~(\ref{Q1}) and Eq.~(\ref{Q2}) will still hold if one uses
this asymptotic scheme. In the following, this asymptotic scheme is
used in which the relationships, $Q_{2}=0$ and $W=Q_{1}$, hold.

\begin{figure*}
\includegraphics[width=14cm]{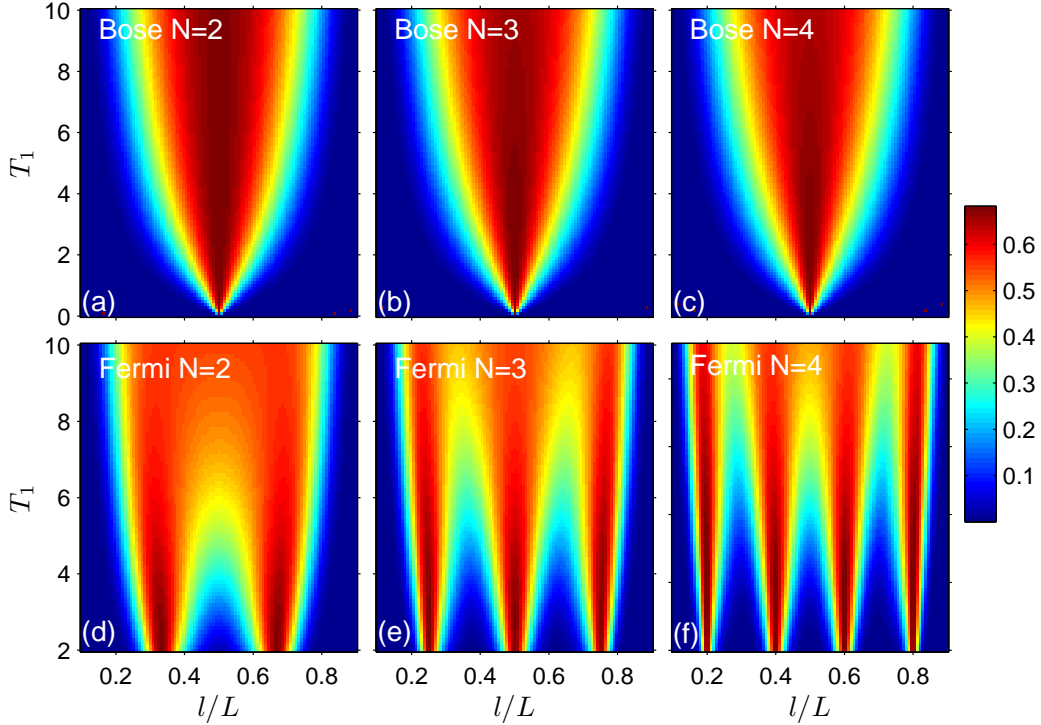} \caption{\label{fig2} (Color Online) $W/T_{1}$ as the function of the insertion
position $l/L$ and the working temperature $T_{1}$. (1) A Bose system
with $N=2$. (2) A Bose system with $N=3$. (3) A Bose system with
$N=4$. (4) A Fermi system with $N=2$. (5) A Fermi system with $N=3$.
(6) A Fermi system with $N=4$.}
\end{figure*}

Since the initial state of the demon is a pure state, which corresponds
to the conditions $T_{2}=0$ and $p_{j}=\delta_{j0}$, the work extracted
can be given by the relationship $W=Q_{1}$ and (\ref{Q1}) as 
\begin{eqnarray}
W & = & T_{1}\left[H\left(\{P_{i}(l)\}\right)+\sum_{i=0}^{N}P_{i}(l)\ln P_{i}(l_{i})\right]\nonumber \\
 & = & T_{1}\sum_{i=0}^{N}P_{i}(l)\ln\left(\frac{P_{i}(l_{i})}{P_{i}(l)}\right).\label{W'}
\end{eqnarray}
 This is just the same as that in Ref.~\cite{Ueda2010} and it can
be calculated by a different way as in Ref.~\cite{Kim2011_revisit}.
Then we will use Eq.~(\ref{W'}) to study the Bose system and the
Fermi system separately.

First of all, we study the behavior of $W/T_{1}$ when $T_{1}\rightarrow0$.
According to Eq.~(\ref{W'}), one gets 
\begin{equation}
\frac{W}{T_{1}}=\sum_{i=0}^{N}P_{i}(l)\ln\left(\frac{P_{i}(l_{i})}{P_{i}(l)}\right).\label{a3_W/T}
\end{equation}
 Here, the expansion position $l_{i}$ have been optimized as in Sec.~\ref{sec4}.
Thus $P_{i}(l_{i})$ is always no less than $P_{i}(l)$. For a deterministic
distribution, $\{P_{i}(l)\vert P_{i}(l)=\delta_{ij}\}$, $W/T_{1}$
in Eq.~(\ref{a3_W/T}) becomes to be $\ln P_{j}(l_{j})$. Due to
the optimization of $l_{j}$, we have $1\geqslant P_{j}(l_{j})\geqslant P_{j}(l)=1$,
or, $P_{j}(l_{j})=1$. Thus, it is concluded that $W/T_{1}$ is always
$0$ when $\{P_{i}(l)\}$ is a deterministic distribution. In other
words, the necessary condition for a non-zero $W/T_{1}$ is that the
informational entropy $H(\{P_{i}(l)\})$ of the distribution $\{P_{i}(l)\}$
is non-zero.

Using the joint entropy theorem \cite{Nielsen_book}, one gets 
\begin{eqnarray}
S_{s,\mathrm{ins}}=H(\{P_{i}(l)\})+\sum_{i}P_{i}(l)S_{i},
\end{eqnarray}
 where $S_{\mathrm{s,ins}}$ is the entropy of the working substance
after the insertion and $S_{i}$ represents the working substance's
entropy in the condition that the number of the particles in the right
compartment is $i$. Due to the non-negative property of entropy,
one finds $S_{\mathrm{s,ins}}\geqslant H(\{P_{i}(l)\})$. Therefore
if the entropy of the working substance is zero, one will find that
$H(\{P_{i}(l)\})=0$ and $W/T_{1}=0$. For a system with its ground
state non-degenerate, its entropy always tends to zero when temperature
tends to zero, which is referred to as the third law of thermodynamics.
Thus, it is found that, if the insertion position $l$ makes the ground
state of the working substance non-degenerate, one will always have
\begin{equation}
\lim_{T_{1}\rightarrow0}\frac{W}{T_{1}}=0.
\end{equation}
 However, if the insertion position $l$ makes the ground state of
the working substance degenerate, there will be a chance that $W/T_{1}$
tends to non-zero. Such a position $l$ is called a degenerate point.
Thus, the relationship between the low-temperature behavior of $W/T_{1}$
and the degenerate point has been established. The similar result
is found in Ref.~\cite{Kim2011_law3}.

Then, we look for the degenerate points for both the ideal Bose system
and the ideal Fermi system. For the ideal Bose system, all particles
condense in the ground state of the system when temperature is zero.
Thus the only chance for the ground state to be degenerate is that
the energy of the left compartment's single-particle ground state
is the same as that of the right compartment, which happens when the
widths of the two compartments are equal to each other. Thereafter
it is found that there is only one degenerate point for the ideal
Bose system, i.e., $l=L/2$.

\begin{figure}[!htb]
\includegraphics[width=8cm]{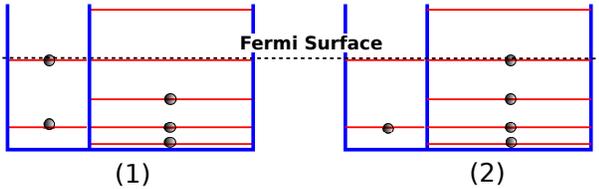} \caption{\label{fermi_degen} (Color Online) Degenerate ground states for a
Fermi gas. When the Fermi surface contains more than one single-particle
states, there is a chance that the ground state of the Fermi gas is
degenerate. For example, the states in (1) and (2) are two ground
states of the system. }
\end{figure}

For a Fermi system, the particles satisfy the Pauli's exclusion principle.
When temperature is zero, all the energy levels under the Fermi surface
are occupied. Thus if the $i$-th single-particle energy level of
the left compartment and the $j$-th single-particle energy level
of the left compartment are both on the Fermi surface, the ground
state of the total gas may be degenerate, see Fig.~\ref{fermi_degen}.
The index $i$ and $j$ should satisfy the relationship that $i+j-1$
equals to the number of the particles $N$. Thus there are $N$ situations
corresponding to $i=1,2,\cdots,N$. Thereafter it is found that there are $N$
degenerate points for an ideal Fermi system with $N$ particles. As
the two compartments are modeled as two one-dimensional infinite square
wells with widths $l$ and $L-l$ respectively, one can workout the
spectrum of them as in Eq.~(\ref{spectrum}). Thus the conditions
of the degenerate points can be written as 
\begin{equation}
\frac{\left(\hbar i\pi\right)^{2}}{2ml^{2}}=\frac{\left(\hbar(N+1-i)\pi\right)^{2}}{2m(L-l)^{2}}.
\end{equation}
 Therefore the positions of the Fermi system's degenerate points are
\begin{equation}
\frac{l}{L}=\frac{i}{N+1}\mbox{ for \ensuremath{i=1,2,\cdots,N}}.
\end{equation}
So far we have revealed the qualitative difference of the work extracted
by the engine between the Bose type and the Fermi type working substance
at low temperature. That difference in the quantum statistical behavior
between the Bose and the Fermi working substance will disappear as
the temperature $T_{1}$ increase. The numerical simulation in Fig.~\ref{fig2}
quantitively illustrates the statistical behavior at finite temperature.
When $T_{1}$ tends to zero, the numerical calculations confirm the
above results. Similar result is also found in Ref.~\cite{Kim2011_law3}.

\section{\label{sec6} Conclusion}

In summary, the multi-particle SHE has been studied in a complete
quantum framework, where the re-initialization process of the demon
is discussed in detail. The errors are introduced in the demon so
that one can use a finite-temperature heat sink to cool down the demon
while the traditional demon must be cooled down by a zero-temperature
heat sink. For a single-particle SHE, there is a scheme that the whole
cycle of the SHE can be revisible. We illustrated that the efficiency
of such a scheme is indeed the efficiency of Carnot cycle's and other
schemes have lower efficiency. And it is proved that SHE can not exceed
the limit of the SLoT. 

On the other hand, the properties of the SHE with Bose or Fermi particles
are studied. In the high-temperature regime, these two distributions
make no difference. Both of them reduce to the result of Maxwell distribution.
In contrast, the Bose system and the Fermi system differ from each
other in the low-temperature regime. The origin of the difference
is the difference of the degenerate points of these two distributions.
The SHE whose insertion position is a degenerate point extracts more
work in a cycle. If the temperature is absolute zero, the ratio of
the work extracted to the working temperature will show some uncontinuous
properties.

This work was supported by NSFC through grants 10974209 and 10935010
and by the National 973 program (Grant No. 2006CB921205).

\appendix

\section{\label{appendix1}Multi-Particle Partition Function}

In this appendix, the explicit expressions of the multi-particle partition
functions for the ideal Bose gas and the ideal Fermi gas are presented.
Let $Z_{n}(l,\beta)$
be the canonical partition function for $n$ particles, where $\beta$
represents the inverse temperature of the system and $l$ represents
some parameters of the system, e.g., in our paper $l$ means the width
of the infinite square well. To present the explicit expression of
$Z_{n}(l,\beta)$, the explicit expression of the single-particle
partition function $Z_{1}(l,\beta)$ is given firstly. Then the relationship
between the multi-particle partition function and the single-particle
partition function is shown.

The single-particle partition function does not depend on which distribution
the particle satisfies. Once one knows the single-particle spectrum
of the system, one can write down its canonical partition function.
For example, the single-particle spectrum in a one-dimensional infinite
square well with width $l$ is 
\begin{equation}
E_{i}(l)=\frac{\left(\hbar i\pi\right)^{2}}{2Ml^{2}},i=1,2,3,...,\label{spectrum}
\end{equation}
where $M$ is the mass of the particle and $i$ is the index of the
energy levels. Then the single-particle partition function of this
system is 
\begin{equation}
Z_{1}(l,\beta)=\sum_{i=1}^{\infty}e^{-\beta E_{i}(l)}.
\end{equation}

When it comes to multi-particle partition function, we first take the 
two-particle system as an example. In this condition, one gets
\begin{equation}
 Z_{2}^{\mathrm{bose(fermi)}}=
\frac{1}{2}\left[\left(\sum_{m=1}^{\infty}e^{-\beta E_{m}}\right)^{2}
\pm \sum_{m=1}^{\infty}e^{-2\beta E_{m}}\right],
\end{equation}
where the indices $\beta$ and $l$ have been omitted for simplicity.
Here, the summation, $\sum_{m=1}^{\infty}\exp(-2\beta E_{m})$, can
be viewed as a single-particle partition function with spectrum $\{2E_{m}\}$.
Thus, it is found
that $Z_{2}^{\mathrm{bose(fermi)}}$ can be decomposed into 
some single-particle partition functions. 
In general, all the partition functions of the ideal gas,
$Z_{n}^{\mathrm{bose(fermi)}}$, can be decomposed
into some single-particle partition functions by this ways.
The next task is to workout the coefficients in the decomposition.

In the following discussion, the system considered has discrete spectrum
and the index of the energy levels is denoted by $i=1,2,3,\cdots$.
Let $n_{i}$ be the number of the particles in the $i$-th level. 
To be convenient, we denote $\exp(-\beta E_{i})$ as $x_i$ for $i=1,2,\cdots$ and
the multi-particle partition function becomes to be
\begin{equation}
 Z_n=\sum_{\{n_{i}\}} \prod_{i=1}^{\infty} x_i^{n_i}.
\end{equation}
Here, the summation condition is that $\sum_i n_i=n$ and $n_i=0,1,2,\cdots$ for Bose 
case or $n_i=0,1$ for Fermi case. It is not difficult to show that
\begin{eqnarray}
 H(t) &\triangleq& \sum_{n=0}^{\infty} Z_n^{\mathrm{bose}} t^n=\prod_i \frac{1}{1-x_i t},\\
 E(t) &\triangleq& \sum_{n=0}^{\infty} Z_n^{\mathrm{fermi}} t^n=\prod_i (1+x_i t).
\end{eqnarray}
In fact, if $t=\exp(-\beta\mu)$, one will find that $H(t)$ and $E(t)$ are
nothing but the grand canonical partition functions for ideal Bose
case and ideal Fermi case respectively. In mathematics, $H(t)$ and $E(t)$
are the generating functions for $Z_n^{\mathrm{bose}}$ and $Z_n^{\mathrm{fermi}}$
respectively and they have the following simple relationship
\begin{equation}
 H(t)=\frac{1}{E(-t)} \label{H_E}.
\end{equation}
Next, the single-particle partition function with spectrum $\{jE_{m}\}$,
$\sum_{m=1}^{\infty}\exp(-j\beta E_{m})=\sum_i x_i^j$, is denoted 
as $P_j$ for $j=1,2,\cdots$.
Its generating function is
\begin{eqnarray}
 P(t) &\triangleq& \sum_{j=1}^{\infty} P_j t^j=\sum_{j=1}^{\infty} \sum_i (x_i t)^j \nonumber\\
&=& \sum_i \sum_{j=1}^{\infty} (x_i t)^j=\sum_i \frac{x_i t}{1-x_i t}.
\end{eqnarray}
Thus, it is found that
\begin{equation}
 P(t)=\frac{t}{H(t)} \frac{\mathrm{d}H(t)}{\mathrm{d} t},
\end{equation}
i.e., 
\begin{equation}
 \frac{P(t)}{t}=\frac{\mathrm{d}}{\mathrm{d}t}\mathrm{ln}H(t).
\end{equation}
By indefinite integration, one finds that
\begin{equation}
 \sum_{j=1}^{\infty} \frac{1}{j} P_j t^j=\mathrm{ln}H(t)+C,
\end{equation}
where $C$ is the constant of integration and it is not difficult
to find that $C=0$. So, there is the following important relationship
\begin{equation}
 H(t)=\exp(\sum_{j=1}^{\infty} \frac{1}{j} P_j t^j). \label{H(t)}
\end{equation}
Compare the coefficient of $t^n$ in both the two sides of Eq.(\ref{H(t)})
and one can find that
\begin{equation}
 Z_n^{\mathrm{bose}}=\sum_{\{i_\alpha\}} \frac{1}{z(\{i_\alpha\})} P(\{i_\alpha\}),
\end{equation}
where $\{i_\alpha\}=(i_1,i_2,\cdots,i_n)$ and the summation condition is
$\sum_\alpha \alpha i_\alpha=n$ which corresponds a summation for all
the Young diagrams with $n$ boxes. The $z(\{i_\alpha\})$ is defined as
\begin{equation}
 z(\{i_\alpha\})=i_1 !1^{i_1}\cdot i_2 !2^{i_2} \cdots i_n !n^{i_n},
\end{equation}
and the $P(\{i_\alpha\})$ is the product of some single-particle partition
functions as
\begin{equation}
 P(\{i_\alpha\})=P_1^{i_1}\cdot P_2^{i_2} \cdots P_n^{i_n}.
\end{equation}
Thus we have decomposed the multi-particle partition function into some 
single-particle partition functions for Bose case. For Fermi case, using
the relationship in Eq.(\ref{H_E}), one finds that
\begin{equation}
 E(t)=\exp(\sum_{j=1}^{\infty} \frac{(-1)^{j+1}}{j} P_j t^j). \label{E(t)}
\end{equation}
Also, compare the coefficient of $t^n$ in both the two sides of Eq.(\ref{E(t)})
and it is found that
\begin{equation}
 Z_n^{\mathrm{fermi}}=
\sum_{\{i_\alpha\}} \frac{(-1)^{\sum_\alpha (i_{\alpha}-1)}}{z(\{i_\alpha\})}
P(\{i_\alpha\}),
\end{equation}
where the summation condition is the same as in the Bose case. 

\section{\label{appendix2}The Upper Limit of $Q_{1}$}

In this appendix, the upper limit of $Q_{1}$ in Eq.~(\ref{Q1})
is estimated. To this end, a two-particle SHE is used as an example
firstly, in which the demon considered is of three-level. The function
$f_{i}(j)$ is chosen as $f_{i}(j)\equiv i+j\pmod{3}$. Thus the heat
absorbed from the heat resource is the one shown in Eq.~(\ref{example_eff}).
To estimate the upper limit of $Q_{1}$, Eq.~(\ref{example_eff})
should be re-written as \begin{widetext} 
\begin{eqnarray}
Q_{1} & =T_{1}H(\{P_{i}(l)\}) & +P_{0}(l)p_{0}\ln P_{0,0}+P_{1}(l)p_{2}\ln P_{1,0}+P_{2}(l)p_{1}\ln P_{2,0}\nonumber \\
 &  & +P_{0}(l)p_{1}\ln P_{0,1}+P_{1}(l)p_{0}\ln P_{1,1}+P_{2}(l)p_{2}\ln P_{2,1}\nonumber \\
 &  & +P_{0}(l)p_{2}\ln P_{0,2}+P_{1}(l)p_{1}\ln P_{1,2}+P_{2}(l)p_{0}\ln P_{2,2},\label{B1}
\end{eqnarray}
 \end{widetext} where $P_{i,j}$ satisfy the constraints $P_{i,j}=P_{i}(l_{j})$
for all $i$ and $j$. Thus the original optimization problem becomes
to be an optimization problem with constraints. This optimization
problem with constraints has a maximum solution $Q_{1\mathrm{max}}$
since $Q_{1}$ in Eq.~(\ref{B1}) has an upper bound $T_{1}H(\{P_{i}(l)\})$
and the domain of $\{P_{i,j}\}$ is closed. Substitute these constraints
by the weaker constraints $\sum_{i}P_{i,j}=1$ for all $j$, then
we get another optimization problem with constraints. After solving
this problem we will get another maximum $Q_{1\mathrm{max}}^{\prime}$.
Because the constraints in the latter problem are weaker than the
constraints in the former one, we have $Q_{1\mathrm{max}}\leqslant Q_{1\mathrm{max}}^{\prime}$.
Thus $Q_{1\mathrm{max}}^{\prime}$ is a upper limit of $Q_{1}$. The
calculation of $Q_{1\mathrm{max}}^{\prime}$ can be performed in terms
of Lagrange multipliers. Namely, one only needs to solve an optimization
problem without constraint in which the target function is 
\begin{equation}
Q_{1}(\{P_{i,j}\})+\lambda_{0}\sum_{i}P_{i,1}+\lambda_{1}\sum_{i}P_{i,0}+\lambda_{2}\sum_{i}P_{i,2},
\end{equation}
 where $\lambda_{0}$, $\lambda_{1}$ and $\lambda_{2}$ are three
Lagrange multipliers. Solving this constraint-free optimization problem,
one finds that the maximum solution $Q_{1\mathrm{max}}^{\prime}$
is reached when $P_{i,j}$ satisfies the following relationships 
\begin{eqnarray}
P_{0,0}:P_{1,0}:P_{2,0} & = & P_{0}(l)p_{0}:P_{1}(l)p_{2}:P_{2}(l)p_{1}\nonumber \\
P_{0,1}:P_{1,1}:P_{2,1} & = & P_{0}(l)p_{1}:P_{1}(l)p_{0}:P_{2}(l)p_{2}\\
P_{0,2}:P_{1,2}:P_{2,2} & = & P_{0}(l)p_{2}:P_{1}(l)p_{1}:P_{2}(l)p_{0}.\nonumber 
\end{eqnarray}

For a general situation, this trick can also be used to find an upper
limit of the heat absorbed from the heat resource $Q_{1}$. To this
end, one only needs to maximize $Q_{1}$ under the constraints $\sum_{i}P_{i}(l_{j})=1$
for all $j$. Thus the upper limit of $Q_{1}$ is given as \begin{widetext}
\begin{eqnarray}
Q_{1} & \leqslant & T_{1}\left[H\left(\{P_{i}(l)\}\right)+\sum_{i=0}^{N}\sum_{j=0}^{N}P_{i}(l)p_{f_{i}^{-1}(j)}\ln\frac{P_{i}(l)p_{f_{i}^{-1}(j)}}{\sum_{i=0}^{N}P_{i}(l)p_{f_{i}^{-1}(j)}}\right]\nonumber \\
 & = & T_{1}\left[H\left(\{\sum_{i=0}^{N}P_{i}(l)p_{f_{i}^{-1}(j)}\}\right)-H\left(\{p_{i}\}\right)\right].
\end{eqnarray}
 \end{widetext}


\begin{thebibliography}{References}
\bibitem{MD_book} \textsl{Maxwell's Demon 2: Entropy, Classical and
Quantum Information, Computing}, edited by H.~S.~Leff and A.~F.~Rex
(Institute of Physics, Bristol, 2003).

\bibitem{Szilard1929} L.~Szilard, Z.~Phys. \textbf{53}, 840 (1929).

\bibitem{Brillouin1951} L.~Brillouin, J. Appl. Phys. \textbf{22},
334 (1951).

\bibitem{Bennett1982} C.~H.~Bennett, Int. J. Theor. Phys. \textbf{21},
905 (1982); Sci. Am. \textbf{257}, 108 (1987).

\bibitem{Landauer1961} R.~Landauer, IBM J. Res. Dev. \textbf{5},
183 (1961).

\bibitem{Ueda2010} S.~W.~Kim, T.~Sagawa, S.~D.~Liberato, M.~Ueda,
Phys. Rev. Lett. \textbf{106}, 070401(2011).

\bibitem{Quan2006} H.~T.~Quan, Y.~D.~Wang, Yu-xi~Liu, C.~P.~Sun,
and F.~Nori, Phys. Rev. Lett. \textbf{97}, 180402 (2006).

\bibitem{Dong2011} H.~Dong, D.~Z.~Xu, C.~Y.~Cai, and C.~P.~Sun,
Phys. Rev. E \textbf{83}, 061108 (2011).

\bibitem{Nori2009} K.~Maruyama, F.~Nori, and V.~Vedral, Rev. Mod.
Phys. \textbf{81}, 1 (2009).

\bibitem{Plenio and Vitelli 2001} M.~B.~Plenio and V.~Vitelli,
Contemp. Phys. \textbf{42}, 25 (2001).

\bibitem{Lloyd1997} S.~Lloyd, Phys. Rev. A \textbf{56}, 3374 (1997).

\bibitem{Kim2011_revisit} K.~H.~Kim and S.~W.~Kim, Phys. Rev.
E \textbf{84}, 012101 (2011).

\bibitem{Kim2011_law3} K.~H.~Kim and S.~W.~Kim, arXiv:1108.3644v2.

\bibitem{Nielsen_book} M.~A.~Nielsen and I.~L.~Chuang, \textsl{Quantum
Computation and Quantum Information}, (Cambridge University Press,
London, 2000)\end{thebibliography}
\end{document}